\documentclass{article}

\usepackage{natbib}

\usepackage{graphicx}

\usepackage{color} 

\usepackage{amssymb}

\usepackage{times}

\begin{document}


\title{Monte-Carlo Simulations of the First Passage Time for %
Multivariate Jump-Diffusion Processes in Financial Applications}

\author{DI ZHANG and RODERICK V.N. MELNIK
\footnote{Corresponding author. Email: rmelnik@wlu.ca, Tel.: +1 519
8841970, Fax: +1 519 8869351.}}

\date{Mathematical Modelling and Computational Sciences, Wilfrid
Laurier University, Waterloo, ON, Canada N2L 3C5}

\maketitle

\begin{abstract}
\noindent Many problems in finance require the information on the
first passage time (FPT) of a stochastic process. Mathematically,
such problems are often reduced to the evaluation of the probability
density of the time for such a process to cross a certain level, a
boundary, or to enter a certain region. While in other areas of
applications the FPT problem can often be solved analytically, in
finance we usually have to resort to the application of numerical
procedures, in particular when we deal with jump-diffusion
stochastic processes (JDP). In this paper, we propose a
Monte-Carlo-based methodology for the solution of the first passage
time problem in the context of multivariate (and correlated)
jump-diffusion processes. The developed technique provide an
efficient tool for a number of applications, including credit risk
and option pricing. We demonstrate its applicability to the analysis
of the default rates and default correlations of several different,
but correlated firms via a set of empirical data.

\vspace*{2ex}\noindent\textit{Keywords}: First passage time; Monte
Carlo simulation; Multivariate jump-diffusion processes; Credit risk

\end{abstract}



\section{Introduction}
\label{Introduction}

Credit risk can be defined as the possibility of a loss occurring
due to the financial failure to meet contractual debt obligations.
This is one of the measures of the likelihood that a party will
default on a financial agreement. There exist two classes of credit
risk models \citep{Zhou:2001:jump}, structural models and reduced
form models. Structural models can be traced back to the influential
works by Black, Scholes and Merton
\citep{Black-Scholes:1973,Merton:1973}, while reduced form models
seem to originate from contributions by \citet{Jarrow:1997}. The
major focus in this contribution is given to structural credit-risk
models.

In structural credit-risk models, a default occurs when a company
cannot meet its financial obligations, or in other words, when the
firm's value falls below a certain threshold. One of the major
problems in the credit risk analysis is when a default occurs within
a given time horizon and what is the default rate during such a time
horizon. This problem can be reduced to a first passage time (FPT)
problem, that can be formulated mathematically as a certain
stochastic differential equation (SDE). It concerns with the
estimation of the probability density of the time for a random
process to cross a specified threshold level.

An important phenomenon that we account for in our discussion lies
with the fact that, in the market economy, individual companies are
inevitably linked together via dynamically changing economic
conditions. Therefore, the default events of companies are often
correlated, especially in the same industry. \citet{Zhou:2001:corr}
and \citet{Hull:2001} were the first to incorporate default
correlation into the Black-Cox first passage structural model, but
they have not included the jumps. As pointed out in
\citet{Zhou:2001:jump,Kou:2003}, the standard Brownian motion model
for market behavior falls short of explaining empirical observations
of market returns and their underlying derivative prices. In the
meantime, jump-diffusion processes (JDPs) have established
themselves as a sound alternative to the standard Brownian motion
model \citep{Atiya:2005}. Multivariate jump-diffusion models provide
a convenient framework for investigating default correlation with
jumps and become more readily accepted in the financial world as an
efficient modeling tool.

However, as soon as jumps are incorporated in the model, except for
very basic applications where analytical solutions are available,
for most practical cases we have to resort to numerical procedures.
Examples of known analytical solutions include problems where the
jump sizes are doubly exponential or exponentially distributed
\citep{Kou:2003} as well as the jumps can have only nonnegative
values (assuming that the crossing boundary is below the process
starting value) \citep{Blake:1973}. For other situations, Monte
Carlo methods remain a primary candidate for applications.

The conventional Monte Carlo methods are very straightforward to
implement. We discretize the time horizon into $N$ intervals with
$N$ being large enough in order to avoid discretization bias
\citep{Platen:2003}. The main drawback of this procedure is that we
need to evaluate the processes at each discretized time which is
very time-consuming. Many researchers have contributed to the field
of enhancement of the efficiency of Monte Carlo simulations. Among
others, \citet{Platen:2002} discussed the solution of SDEs in the
framework of weak discrete time approximations and
\citet{Platen:2006} considered the strong approximation where the
SDE is driven by a high intensity Poisson process. Atiya and
Metwally \citep{Atiya:2005,Metwally:2002} have recently developed a
fast Monte Carlo-type numerical methods to solve the FPT problem. In
our recent contribution, we reported an extension of this fast
Monte-Carlo-type method in the context of multiple non-correlated
jump-diffusion processes \citep{Zhang:2006}.

In this contribution, we generalize our previous fast Monte-Carlo
method (for non-correlated jump-diffusion cases) to multivariate
(and correlated) jump-diffusion processes. The developed technique
provides an efficient tool for a number of applications, including
credit risk and option pricing. We demonstrate the applicability of
this technique to the analysis of the default rates and default
correlations of several different correlated firms via a set of
empirical data.

The paper is organized as follows, Section \ref{Model} provides
details of our model in the context of multivariate jump-diffusion
processes. The algorithms and the calibration of the model are
presented in Section \ref{Methodology}. Section \ref{Application}
demonstrates how the model works via analyzing the credit risk of
multi-correlated firms. Conclusions are given in Section
\ref{Conclusion}.

\section{Model description}
\label{Model}

As mentioned in the introduction, when we deal with jump-diffusion
stochastic processes, we usually have to resort to the application
of numerical procedures. Although Monte Carlo procedure provide a
natural in such case, the one-dimensional fast Monte-Carlo method
cannot be directly generalized to the multivariate and correlated
jump-diffusion case (e.g. \citet{Zhang:2006}). The difficulties
arise from the fact that the multiple processes as well as their
first passage times are indeed correlated, so the simulation must
reflect the correlations of first passage times. In this
contribution, we propose a solution to circumvent these difficulties
by combining the fast Monte-Carlo method of one-dimensional
jump-diffusion processes and the generation of correlated
multidimensional variates. This approach generalizes our previous
results for the non-correlated jump-diffusion case to multivariate
and correlated jump-diffusion processes.

In this section, first, we present a probabilistic description of
default events and default correlations. Next, we describe the
multivariate jump-diffusion processes and provide details on the
first passage time distribution under the one-dimensional Brownian
bridge (the sum-of-uniforms method which is used to generate
correlated multidimensional variates will be described in Section
\ref{subsection:SOU}). Finally, we presents kernel estimation in the
context of our problem that can be used to represent the first
passage time density function.

\subsection{Default and default correlation}
\label{subsection:default}

In a structural model, a firm $i$ defaults when it can not meet its
financial obligations, or in other words, when the firm assets value
$V_i(t)$ falls below a threshold level $D_{V_i}(t)$. Generally
speaking, finding the threshold level $D_{V_i}(t)$ is one of the
challenges in using the structural methodology in the credit risk
modeling, since in reality firms often rearrange their liability
structure when they have credit problems. In this contribution, we
use an exponential form defining the threshold level
$D_{V_i}(t)=\kappa_i\exp(\gamma_i t)$ as proposed by
\citet{Zhou:2001:corr}, where $\gamma_i$ can be interpreted as the
growth rate of firm's liabilities. Coefficient $\kappa_i$ captures
the liability structure of the firm and is usually defined as a
firm's short-term liability plus 50\% of the firm's long-term
liability. If we set $X_i(t)=\ln[V_i(t)]$, then the threshold of
$X_i(t)$ is $D_i(t)=\gamma_i t+\ln(\kappa_i)$. Our main interest is
in the process $X_i(t)$.

Prior to moving further, we define a default correlation that
measures the strength of the default relationship between different
firms. Take two firms $i$ and $j$ as an example, whose probabilities
of default are $P_i$ and $P_j$, respectively. Then the default
correlation can be defined as
\begin{equation}
  \rho_{ij}=\frac{P_{ij}-P_{i}P_{j}}{\sqrt{P_i(1-P_i)P_j(1-P_j)}},
  \label{Eq:corr}
\end{equation}
where $P_{ij}$ is the probability of joint default.

From Eq. (\ref{Eq:corr}) we have
$P_{ij}=P_{i}P_{j}+\rho_{ij}\sqrt{P_i(1-P_i)P_j(1-P_j)}$. Let us
assume that $P_{i}=P_{j}=5$\%. If these two firms are independent,
i.e., the default correlation $\rho_{ij}=0$, then the probability of
joint default is $P_{ij}=0.25$\%. If the two firms are positively
correlated, for example, $\rho_{ij}=0.4$, then the probability of
both firms default becomes $P_{ij}=2.15$\% that is almost 10 times
higher than in the former case. Thus, the default correlation
$\rho_{ij}$ plays a key role in the joint default with important
implications in the field of credit analysis. \citet{Zhou:2001:corr}
and \citet{Hull:2001} were the first to incorporate default
correlation into the Black-Cox first passage structural model.

\citet{Zhou:2001:corr} has proposed a first passage time model to
describe default correlations of two firms under the ``bivariate
diffusion process'':
\begin{equation}
  \left[\begin{array}{cc}
   X_1(t)\\
   X_2(t)
   \end{array}\right]=
  \left[\begin{array}{cc}
   \mu_1\\
   \mu_2
   \end{array}\right]dt+
   \Omega\left[\begin{array}{cc}
   dz_1\\
   dz_2
   \end{array}\right],
  \label{zhou:two:assets}
\end{equation}
where $\mu_1$ and $\mu_2$ are constant drift terms, $z_1$ and $z_2$
are two independent standard Brownian motions, and $\Omega$ is a
constant $2\times 2$ matrix such that
\[
  \Omega\cdot\Omega'=\left[\begin{array}{cc}
    \sigma_1^2 & \rho\sigma_1\sigma_2 \\
    \rho\sigma_1\sigma_2 & \sigma_2^2\end{array}\right].
\]
The coefficient $\rho$ reflects the correlation between the
movements in the asset values of the two firms. Then the probability
of firm $i$ defaults at time $t$ can be easily calculated as,
\begin{equation}
  P_i(t)=2\cdot N\left(-\frac{X_i(0)-\ln(\kappa_i)}{\sigma_i\sqrt{t}}\right)%
        =2\cdot N\left(-\frac{Z_i}{\sqrt{t}}\right),
  \label{default:zhou:model}
\end{equation}
where
\[
  Z_i\equiv\frac{X_i(0)-\ln(\kappa_i)}{\sigma_i}
  \label{Zi:zhou}
\]
is the standardized distance of firm $i$ to its default point and
$N(\cdot)$ denotes the cumulative probability distribution function
for a standard normal variable.

Furthermore, if we assume that $\mu_i=\gamma_i$, then the
probability of at least one firm defaults by time $t$ can be written
as \citep{Zhou:2001:corr}:
\begin{eqnarray}
  P_{i\cup j}(t)&=&1-\frac{2r_0}{\sqrt{2\pi t}}\cdot\mathrm{e}^{-\frac{r_0^2}{4t}}
    \cdot\sum_{n=1,3,\cdots}\frac{1}{n}\sin\left(\frac{n\pi\theta_0}{\alpha}\right)\nonumber\\
    &&\cdot\left[I_{\frac{1}{2}\left(\frac{n\pi}{\alpha}+1\right)}\left(\frac{r_0^2}{4t}\right)
      +I_{\frac{1}{2}\left(\frac{n\pi}{\alpha}-1\right)}\left(\frac{r_0^2}{4t}\right)\right],
    \label{Eq:Por}
\end{eqnarray}
where $I_{\nu}(z)$ is the modified Bessel function $I$ with order
$\nu$ and
\begin{eqnarray*}
  \alpha&=&\left\{\begin{array}{ll}
    \tan^{-1}\left(-\frac{\sqrt{1-\rho^2}}{\rho}\right),&\mathrm{if}\;\rho<0,\\
    \pi+\tan^{-1}\left(-\frac{\sqrt{1-\rho^2}}{\rho}\right),&\mathrm{otherwise},
    \end{array}\right.\\
  \theta_0&=&\left\{\begin{array}{ll}
    \tan^{-1}\left(\frac{Z_2\sqrt{1-\rho^2}}{Z_1-\rho Z_2}\right),&\mathrm{if}\;(\cdot)>0,\\
    \pi+\tan^{-1}\left(\frac{Z_2\sqrt{1-\rho^2}}{Z_1-\rho Z_2}\right),&\mathrm{otherwise},
    \end{array}\right.\\
  r_0&=&Z_2/\sin(\theta_0).
\end{eqnarray*}

Then, the default correlation of these two firms is
\begin{equation}
  \rho_{ij}(t)=\frac{P_{i}(t)+P_{j}(t)-P_{i}(t)P_{j}(t)-P_{i\cup j}(t)}%
    {\sqrt{P_i(t)[1-P_i(t)]P_j(t)[1-P_j(t)]}}.
  \label{Eq:corr2}
\end{equation}

However, none of the above known models includes jumps in the
processes. At the same time, it is well-known that jumps are a major
factor in the credit risk analysis. With jumps included in such
analysis, a firm can default instantaneously because of a sudden
drop in its value which is impossible under a diffusion process.
\citet{Zhou:2001:jump} has shown the importance of jump risk in
credit risk analysis of an obligor. He implemented a simulation
method to show the effect of jump risk in the credit spread of
defaultable bonds. He showed that the misspecification of stochastic
processes governing the dynamics of firm value, i.e., falsely
specifying a jump-diffusion process as a continuous Brownian motion
process, can substantially understate the credit spreads of
corporate bonds. Therefore, for multiple processes, considering the
simultaneous jumps can be a better way to estimate the correlated
default rates. The multivariate jump-diffusion processes can provide
a convenient way to describe multivariate and correlated processes
with jumps.

\subsection{Multivariate jump-diffusion processes}
\label{subsection:MJD}

Let us consider a complete probability space $(\Omega,F,P)$ with
information filtration $(F_t)$. Suppose that $X_t=\ln(V_t)$ is a
Markov process in some state space $D\subset\mathbb{R}^n$, solving
the stochastic differential equation \citep{Duffie:2000}:
\begin{equation}
  dX_t=\mu(X_t)dt+\sigma(X_t)dW_t+dZ_t,
  \label{AJD:DiffEq}
\end{equation}
where $W$ is an $(F_t)$-standard Brownian motion in $\mathbb{R}^n$;
$\mu:D\rightarrow\mathbb{R}^n$,
$\sigma:D\rightarrow\mathbb{R}^{n\times n}$, and $Z$ is a pure jump
process whose jumps have a fixed probability distribution $\nu$ on
$\mathbb{R}^n$ such that they arrive with intensity
$\{\lambda(X_t):t\ge 0\}$, for some
$\lambda:D\rightarrow[0,\infty)$. Under these conditions, the above
model is reduced to an affine model if
\citep{Ahangarani:2005,Duffie:2000}:
\begin{eqnarray}
  & & \mu(X_t,t) = K_0 + K_1 X_t\nonumber\\
  & & (\sigma(X_t,t)\sigma(X_t,t)^\top)_{ij} = (H_0)_{ij}+(H_1)_{ij}X_j\nonumber\\
  & & \lambda(X_t) = l_0+l_1\cdot X_t,
  \label{Eq:AJD:terms}
\end{eqnarray}
where $K=(K_0,K_1)\in\mathbb{R}^n\times\mathbb{R}^{n\times n}$,
$H=(H_0,H_1)\in\mathbb{R}^{n\times n}\times\mathbb{R}^{n\times
n\times n}$, $l=(l_0,l_1)\in\mathbb{R}^n\times\mathbb{R}^{n\times
n}$.

As we mentioned, one of the major problems in the credit risk
analysis is to estimate the default rate of a firm during a given
time horizon. This problem is reduced to a first passage time
problem. In order to obtain a computable multi-dimensional formulas
of FPT distribution, we need to simplify Eq. (\ref{AJD:DiffEq}) and
(\ref{Eq:AJD:terms}) as follows,
\begin{enumerate}
  \item Each $W_t$ in Eq. (\ref{AJD:DiffEq}) is independent;
  \item $K_1=0$, $H_1=0$ and $l_1=0$ in Eq. (\ref{Eq:AJD:terms}) which means
  the drift term, the diffusion process (Brownian motion) and the
  arrival intensity are independent of the state vector $X_t$;
  \item The distribution of jump-size $Z_t$ is also independent with respect
  to $X_t$.
\end{enumerate}

In this scenario, we can rewrite Eq. (\ref{AJD:DiffEq}) as
\begin{equation}
  dX_t=\mu dt+\sigma dW_t+dZ_t,
  \label{JDP:multi}
\end{equation}
where
\[
  \mu = K_0,\;\sigma\sigma^\top = H_0,\;\lambda = l_0.
\]

\subsection{First passage time distribution of Brownian bridge}
\label{subsection:FPTD}

Although for jump-diffusion processes, the closed form solutions are
usually unavailable, yet between each two jumps the process is a
Brownian bridge for univariate jump-diffusion process.
\citet{Atiya:2005} have deduced one-dimensional first passage time
distribution in time horizon $[0,T]$. In order to evaluate multiple
processes, we obtain multi-dimensional formulas from Eq.
(\ref{JDP:multi}) and reduce them to computable forms.

First, let us consider a firm $i$, as described by Eq.
(\ref{JDP:multi}), such that its state vector $X_i$ satisfies the
following SDE:
\begin{eqnarray}
  dX_i & = & \mu_{i}dt+\sum_{j}\sigma_{ij}dW_j+dZ_i\nonumber\\
       & = & \mu_{i}dt+\sigma_{i}dW_i+dZ_i,
  \label{JDP:one}
\end{eqnarray}
where $W_i$ is a standard Brownian motion and $\sigma_{i}$ is:
\[
  \sigma_{i}=\sqrt{\sum_{j}\sigma_{ij}^2}.
\]

We assume that in the interval $[0,T]$, total number of jumps for
firm $i$ is $M_i$. Let the jump instants be $T_{1},
T_{2},\cdots,T_{M_i}$. Let $T_{0}=0$ and $T_{M_i+1}=T$. The
quantities $\tau_j$ equal to interjump times, which is
$T_{j}-T_{j-1}$. Following the notation of \citet{Atiya:2005}, let
$X_{i}(T_{j}^{-})$ be the process value immediately before the $j$th
jump, and $X_{i}(T_{j}^{+})$ be the process value immediately after
the $j$th jump. The jump-size is
$X_{i}(T_{j}^{+})-X_{i}(T_{j}^{-})$, and we can use such jump-sizes
to generate $X_{i}(T_{j}^{+})$ sequentially.

If we define $A_i(t)$ as the event consisting of process crossed the
threshold level $D_i(t)$ for the first time in the interval
$[t,t+dt]$, then we have
\begin{equation}
  g_{ij}(t)=p(A_i(t)\in dt|X_i(T_{j-1}^{+}),X_i(T_{j}^{-})).
\end{equation}

If we only consider one interval $[T_{j-1},T_{j}]$, we can obtain
\begin{eqnarray}
  g_{ij}(t) & = &
  \frac{X_i(T_{j-1}^{+})-D_{i}(t)}{2y_i\pi\sigma_{i}^{2}}(t-T_{j-1})^{-\frac{3}{2}}(T_{j}-t)^{-\frac{1}{2}}\nonumber\\
  & & *\exp\left(-\frac{[X_i(T_{j}^{-})-D_{i}(t)-\mu_{i}(T_{j}-t)]^{2}}{2(T_{j}-t)\sigma_{i}^{2}}\right)\nonumber\\
  & & *\exp\left(-\frac{[X_i(T_{j-1}^{+})-D_{i}(t)+\mu_{i}(t-T_{j-1})]^{2}}{2(t-T_{j-1})\sigma_{i}^{2}}\right),
  \label{FPTD:condition}
\end{eqnarray}
where
\[
  y_i=\frac{1}{\sigma_{i}\sqrt{2\pi\tau_{j}}}
    \exp\left(-\frac{[X_i(T_{j-1}^{+})-X_i(T_{j}^{-})+\mu_{i}\tau_{j}]^{2}}{2\tau_{j}\sigma_{i}^{2}}\right).
\]

After getting result in one interval, we combine the results to
obtain the density for the whole interval $[0,T]$. Let $B(s)$ be a
Brownian bridge in the interval $[T_{j-1},T_{j}]$ with
$B(T_{j-1}^{+})=X_i(T_{j-1}^{+})$ and $B(T_{j}^{-})=X_i(T_{j}^{-})$.
Then the probability that the minimum of $B(s_i)$ is always above
the boundary level is
\begin{eqnarray}
  P_{ij} & = & P\left(\inf_{T_{j-1}\leq s_i\leq T_{j}}B(s_i)>D_{i}(t)|B(T_{j-1}^{+})=X_i(T_{j-1}^{+}),B(T_{j}^{-})=X_i(T_{j}^{-})\right)\nonumber\\
   & = & \left\{
     \begin{array}{ll}
       1-\exp\left(-\frac{2[X_i(T_{j-1}^{+})-D_{i}(t)][X_i(T_{j}^{-})-D_{i}(t)]}{\tau_{j}\sigma_{i}^{2}}\right), & \mathrm{if}\;X_i(T_{j}^{-})>D_{i}(t),\\
       0, & \mathrm{otherwise}.
     \end{array}\right.
   \label{BM:default}
\end{eqnarray}

This implies that $B(s_i)$ is below the threshold level, which means
the default happens or already happened, and its probability is
$1-P_{ij}$. Let $L(s_i)\equiv L_i$ denote the index of the interjump
period in which the time $s_i$ (first passage time) falls in
$[T_{L_i-1},T_{L_i}]$. Also, let $I_i$ represent the index of the
first jump, which happened in the simulated jump instant,
\begin{eqnarray}
  I_i & = & \min(j:X_i(T_{k}^{-})>D_{i}(t);k=1,\ldots,j,\;\mathrm{and}\nonumber\\
    &   & \;\;\;\;\;\;\;\;\;\;\;\:X_i(T_{k}^{+})>D_{i}(t);k=1,\ldots,j-1,\;\mathrm{and}
          \;X_i(T_{j}^{+})\leq D_{i}(t)).
  \label{index_first_jump}
\end{eqnarray}

If no such $I_i$ exists, then we set $I_i=0$.

By combining Eq. (\ref{FPTD:condition}), (\ref{BM:default}) and
(\ref{index_first_jump}), we get the probability of $X_i$ crossing
the boundary level in the whole interval $[0,T]$ as
\begin{eqnarray}
  & & P(A_i(s_i)\in ds|X_i(T_{j-1}^{+}),X_i(T_{j}^{-}),j=1,\ldots,M_{i}+1)\nonumber\\
   & = & \left\{
     \begin{array}{ll}
       g_{iL_i}(s_i)\prod_{k=1}^{L_i-1}P_{ik} & \mathrm{if}\;L_i<I_i\;\mathrm{or}\;I_i=0,\\
       g_{iL_i}(s_i)\prod_{k=1}^{L_i-1}P_{ik}+\prod_{k=1}^{L_i}P_{ik}\delta(s_i-T_{I_i}) & \mathrm{if}\;L_i=I_i,\\
       0 & \mathrm{if}\;L_i>I_i,
     \end{array}\right.
\end{eqnarray}
where $\delta$ is the Dirac's delta function.

\subsection{The kernel estimator}
\label{subsection:estimation}

For firm $i$, after generating a series of first passage times
$s_i$, we use a kernel density estimator with Gaussian kernel to
estimate the first passage time density (FPTD) $f$. The kernel
density estimator is based on centering a kernel function of a
bandwidth as follows:
\begin{equation}
  \widehat{f}=\frac{1}{N}\sum_{i=1}^{N}K(h,t-s_{i}),
  \label{Eq:estimator}
\end{equation}
where
\[
  K(h,t-s_{i})=\frac{1}{\sqrt{\pi/2}h}\exp\left(-\frac{(t-s_{i})^{2}}{h^2/2}\right).
\]

The optimal bandwidth in the kernel function $K$ can be calculated
as \citep{Silverman:1986}:
\begin{equation}
  h_{opt}=\left(2N\sqrt{\pi}\int_{-\infty}^{\infty}(f_{t}'')^{2}dt\right)^{-0.2},
  \label{estamate:hopt}
\end{equation}
where $N$ is the number of generated points and $f_{t}$ is the true
density. Here we use the approximation for the distribution as a
gamma distribution, proposed by \citet{Atiya:2005}:
\begin{equation}
  f_{t}=\frac{\alpha^{\beta}}{\Gamma(\beta)}t^{\beta-1}\exp(-\alpha t).
\end{equation}

So the integral in Eq. (\ref{estamate:hopt}) becomes:
\begin{equation}
  \int_{0}^{\infty}(f_{t}'')^{2}dt=
    \sum_{i=1}^{5}\frac{W_{i}\alpha_{i}\Gamma(2\beta-i)}{2^{(2\beta-i)}(\Gamma(\beta))^{2}},
  \label{Eq:hopt2}
\end{equation}
where
\[
  W_{1}=A^{2},\;\;W_{2}=2AB,\;\;W_{3}=B^{2}+2AC,\;\;W_{4}=2BC,\;\;W_{5}=C^{2},
\]
and
\[
  A=\alpha^{2},\;\;B=-2\alpha(\beta-1),\;\;C=(\beta-1)(\beta-2).
\]

From Eq. (\ref{Eq:hopt2}), it follows that in order to get a nonzero
bandwidth, we have to have constraint $\beta$ to be at least equal
to 3.

The kernel estimator for the multivariate case involves the
evaluation of joint conditional interjump first passage time
density, as discussed in Section \ref{Methodology}. The methodology
for such an evaluation is quite involved compared to the
one-dimensional case and we will focus on these details elsewhere.
In what follows we highlight the main steps of the procedure.

\section{The methodology of solution}
\label{Methodology}

\begin{figure}[hbtp]
  \centering
  \includegraphics[width=10cm]{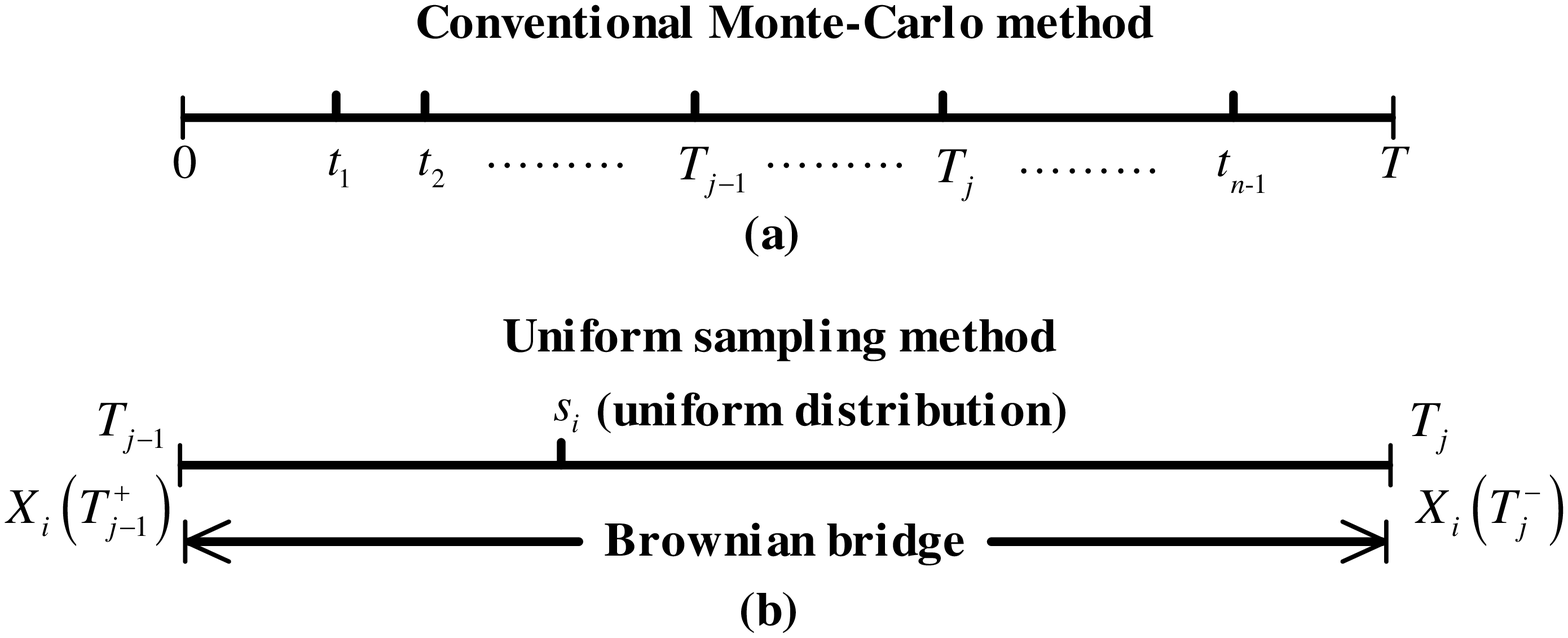}
  \caption{Schematic diagram of (a) conventional Monte Carlo and
           (b) uniform sampling (UNIF) method.}
  \label{Fig:Method}
\end{figure}

First, let us recall the conventional Monte-Carlo procedure in
application to the analysis of the evolution of firm $X_i$ within
the time horizon $[0,T]$. We divide the time horizon into $n$ small
intervals $[0,t_1]$, $[t_1,t_2]$, $\cdots$, $[t_{n-1},T]$ as
displayed in Fig. \ref{Fig:Method}(a). In each Monte Carlo run, we
need to calculate the value of $X_i$ at each discretized time $t$.
As usual, in order to exclude discretization bias, the number $n$
must be large. This procedure exhibits substantial computational
difficulties when applied to jump-diffusion processes. Indeed, for a
typical jump-diffusion process, as shown in Fig.
\ref{Fig:Method}(a), let $T_{j-1}$ and $T_j$ be any successive jump
instants, as described above. Then, in the conventional Monte Carlo
method, although there is no jump occurring in the interval
$[T_{j-1},T_j]$, yet we need to evaluate $X_i$ at each discretized
time $t$ in $[T_{j-1},T_j]$. This very time-consuming procedure
results in a serious shortcoming of the conventional Monte-Carlo
methodology.

To remedy the situation, two modifications of the conventional
procedure were recently proposed \citep{Atiya:2005,Metwally:2002}
that allow us a potential speed-up of the conventional methodology
in 10-30 times. One of the modifications, the uniform sampling
method, involves samplings using the uniform distribution. The other
is the inverse Gaussian density sampling method. Both methodologies
were developed for the univariate case.

The major improvement of the uniform sampling method is based on the
fact that it only evaluates $X_i$ at generated jump times, while
between each two jumps the process is a Brownian bridge (see Fig.
\ref{Fig:Method}(b)). Hence, we just consider the probability of
$X_i$ crossing the threshold in $(T_{j-1},T_j)$ instead of
evaluating $X_i$ at each discretized time $t$. More precisely, in
the uniform sampling method, we assume that the values of
$X_i(T_{j-1}^+)$ and $X_i(T_j^-)$ are known as two end points of the
Brownian bridge, the probability of firm $i$ defaults in
$(T_{j-1},T_j)$ is $1-P_{ij}$ which can be computed according to Eq.
(\ref{BM:default}). Then we generate a variable $s_i$ from a
distribution uniform in the interval
$[T_{j-1},T_{j-1}+\frac{T_{j}-T_{j-1}}{1-P_{ij}}]$. If the generated
point $s_i$ falls in the interjump interval $[T_{j-1},T_j]$, then we
have successfully generated the first passage time $s_i$ and can
neglect the other intervals and perform another Monte Carlo run. On
the other hand, if the generated point $s_i$ falls outside the
interval $[T_{j-1},T_j]$ (which happens with probability $P_{ij}$),
then that point is ``rejected''. This means that no boundary
crossing has occurred in the interval, and we proceed to the next
interval and repeat the whole process again.

Note that the generated $s_i$ is not obtained according to
conditional boundary crossing density $g_{ij}(s_i)$ as described by
Eq. (\ref{FPTD:condition}). In order to obtain an appropriate
density estimate, \citet{Atiya:2005} proposed that the right hand
side summation in Eq. (\ref{Eq:estimator}) can be viewed as a finite
sample estimate of the following:
\begin{eqnarray}
  E_{g_{ij}(s_i)}[K(h,t-s_i)]&\equiv&\int_{T_{j-1}}^{T_j}g_{ij}(s_i)K(h,t-s_i)ds_i\nonumber\\
   &=&\left(\frac{T_j-T_{j-1}}{1-P_{ij}}\right)E_{U(s_i)}[g_{ij}(s_i)K(h,t-s_i)],
\end{eqnarray}
where $E_{g_{ij}(s_i)}$ means the expectation of $s_i$, where $s_i$
obeys the density $g_{ij}(s_i)$. $U(s_i)$ is the uniform density in
$[T_{j-1},T_{j-1}+\frac{T_{j}-T_{j-1}}{1-P_{ij}}]$ from which we
sample the point $s_i$. Therefore, we should weight the kernel with
$\left(\frac{T_{j}-T_{j-1}}{1-P_{ij}}\right)g_{ij}(s_i)$ to obtain
an estimate for the true density.

For the multidimensional density estimate we need to evaluate the
joint conditional boundary crossing density. This problem can be
divided into several one-dimensional density estimate subproblems if
the processes are non-correlated \citep{Zhang:2006}. As for the
multivariate correlated processes, the joint density becomes very
complicated and there are usually no analytical solutions for
higher-dimensional processes \citep{Song1:2006,Wise:2004}. We will
not consider this problem in the current contribution.

Instead, we focus on the further development of the uniform sampling
(UNIF) method and extend it to multivariate and correlated
jump-diffusion processes. In order to implement the UNIF method for
our multivariate model as described in Eq. (\ref{JDP:multi}), we
need to consider several points:
\begin{enumerate}
  \item We assume that the arrival rate $\lambda$ for the Poisson jump
process and the distribution of $(T_j-T_{j-1})$ are the same for
each firm. As for  the jump-size, we generate them by a given
distribution which can be different for different firms to reflect
specifics of the jump process for each firm.
  \item We exemplify our description by considering an exponential
distribution (mean value $\mu_T$) for $(T_j-T_{j-1})$ and a normal
distribution (mean value $\mu_J$ and standard deviation $\sigma_J$)
for the jump-size. We can use any other distribution when
appropriate.
  \item An array \texttt{IsDefault} (whose size is
the number of firms denoted by $N_{\mathrm{firm}}$) is used to
indicate whether firm $i$ has defaulted in this Monte Carlo run. If
the firm defaults, then we set \texttt{IsDefault}$(i)=1$, and will
not evaluate it during this Monte Carlo run.
  \item Most importantly, as we have mentioned before, the default events of
firm $i$ are inevitably correlated with other firms, for example
firm $i+1$. The default correlation of firms $i$ and $i+1$ is
described by Eq. (\ref{Eq:corr2}). Hence, firm $i$'s first passage
time $s_i$ is indeed correlated with $s_{i+1}$ -- the first passage
time of firm $i+1$. We must generate several correlated $s_i$ in
each interval $[T_{j-1},T_{j-1}+\frac{T_{j}-T_{j-1}}{1-P_{ij}}]$
which is the key point for multivariate correlated processes.
\end{enumerate}

Note that the assumption based on using the same arrival rate
$\lambda$ and distribution of $(T_j-T_{j-1})$ for different firms
may seem to be quite idealized. One may argue that the arrival rate
$\lambda$ for the Poisson jump process should be different for
different firms, which implies that different firms endure different
jump rates. However, if we consider the real market economy, once a
firm (called firm ``A'') encounter sudden economic hazard, its
correlated firms may also endure the same hazard. Furthermore, it is
common that other firms will help firm ``A'' to pull out, which may
result in a simultaneous jump for them. Therefore, as a first step,
it is reasonable to employ the simultaneous jumps' processes for all
the different firms.

Next, we will give a brief description of the sum-of-uniforms method
which is used to generate correlated uniform random variables,
followed by the description of the multivariate and correlated UNIF
method and the model calibration.

\subsection{Sum-of-uniforms method}
\label{subsection:SOU}

In the above sections, we have reduced the solution of the original
problem to a series of one-dimensional jump-diffusion processes as
described by Eq. (\ref{JDP:one}). The first passage time
distribution in an interval $[T_{j-1},T_{j}]$ (between two
successive jumps) was obtained in section \ref{subsection:FPTD}. As
mentioned, the default events of firm $i$ are inevitably correlated
with other firms, for example firm $i+1$. In this contribution, we
approximate the correlation of $s_i$ and $s_{i+1}$ as the default
correlation of firm s$i$ and $i+1$ by the following formula:
\begin{equation}
  \rho(s_i,s_{i+1})\approx\rho_{i,i+1}(t)=%
    \frac{P_{i}(t)+P_{i+1}(t)-P_{i}(t)P_{i+1}(t)-P_{i\cup i+1}(t)}%
    {\sqrt{P_i(t)[1-P_i(t)]P_{i+1}(t)[1-P_{i+1}(t)]}},
  \label{Eq:corr:FPT}
\end{equation}
where $t$ can be chosen as the midpoint of this interval.

Therefore, we need to generate several correlated $s_i$ in
$[T_{j-1},T_{j-1}+\frac{T_{j}-T_{j-1}}{1-P_{ij}}]$ whose
correlations can be described by Eq. (\ref{Eq:corr:FPT}). Let us
introduce a new variable $b_{ij}=\frac{T_{j}-T_{j-1}}{1-P_{ij}}$,
then we have $s_i=b_{ij}Y_i+T_{j-1}$, where $Y_i$ are uniformly
distributed in $[0,1]$. Moreover, the correlation of $Y_i$ and
$Y_{i+1}$ is given by $\rho(s_i,s_{i+1})$.

Now we can generate the correlated uniform random variables $Y_1,
Y_2,\cdots$ by using the sum-of-uniforms (SOU) method
\citep{Chen:2005,Willemain:1993} in the following steps:
\begin{enumerate}
  \item Generate $Y_1$ from numbers uniformly distributed in $[0,1]$.
  \item For $i=2,3,\cdots$, generate $W_i\sim U(0,c_{i-1,i})$, where
  $U(0,c_{i-1,i})$ denotes a uniform random number over range
  $(0,c_{i-1,i})$. \citet{Chen:2005} has obtained the relationship of parameter
  $c_{i-1,i}$ and the correlation $\rho(s_{i-1},s_{i})$ (abbreviated as $\rho_{i-1,i}$) as follows:
  \[
  \begin{array}{|l|l|l|}
    \hline
    \mathrm{Correlation} & c_{i-1,i}\ge1 & c_{i-1,i}<1\\
    \hline
    \rho_{i-1,i}\ge0 &
    \frac{1}{c_{i-1,i}}-\frac{0.3}{c_{i-1,i}^2} (\rho_{i-1,i}\le0.7) &
    1-0.5c_{i-1,i}^2+0.2c_{i-1,i}^3 (\rho_{i-1,i}\ge0.7)\\
    \rho_{i-1,i}\le0 &
    -\frac{1}{c_{i-1,i}}+\frac{0.3}{c_{i-1,i}^2} (\rho_{i-1,i}\ge-0.7) &
    -1+0.5c_{i-1,i}^2-0.2c_{i-1,i}^3 (\rho_{i-1,i}\le-0.7)\\
    \hline
  \end{array}
  \]
  If $Y_{i-1}$ and $Y_{i}$ are positively correlated, then let
  \[Z_i=Y_{i-1}+W_i.\]
  If $Y_{i-1}$ and $Y_{i}$ are negatively correlated, then let
  \[Z_i=1-Y_{i-1}+W_i.\]

  Let $Y_i=F(Z_i)$, where for $c_{i-1,i}\ge1$,
  \[
    F(Z)=\left\{\begin{array}{ll}
      Z^2/(2c_{i-1,i}), & 0\le Z\le 1,\\
      (2Z-1)/(2c_{i-1,i}), & 1\le Z\le c_{i-1,i},\\
      1-(1+c_{i-1,i}-Z)^2/(2c_{i-1,i}), & c_{i-1,i}\le Z\le 1+c_{i-1,i},
    \end{array}\right.
  \]
  and for $0<c_{i-1,i}\le1$,
  \[
    F(Z)=\left\{\begin{array}{ll}
      Z^2/(2c_{i-1,i}), & 0\le Z\le c_{i-1,i},\\
      (2Z-c_{i-1,i})/2, & c_{i-1,i}\le Z\le 1,\\
      1-(1+c_{i-1,i}-Z)^2/(2c_{i-1,i}), & 1\le Z\le 1+c_{i-1,i}.
    \end{array}\right.
  \]
\end{enumerate}

\subsection{Uniform sampling method}
\label{subsection:UNIF}

In this subsection, we will describe our algorithm for multivariate
jump-diffusion processes, which is an extension of the
one-dimensional case developed earlier by other authors (e.g.
\citet{Atiya:2005,Metwally:2002}).

Consider $N_{\mathrm{firm}}$ firms in the given time horizon
$[0,T]$. First, we generate the jump instant $T_{j}$ by generating
interjump times $(T_{j}-T_{j-1})$ and set all the
\texttt{IsDefault}$(i)=0(i=1,2,\cdots,N_{\mathrm{firm}})$ to
indicate that no firm defaults at first.

From Fig. \ref{Fig:Method}(b) and Eq. (\ref{JDP:one}), we can
conclude that for each process $X_i$ we can make the following
observations:
\begin{enumerate}
  \item If no jump occurs, as described by Eq. (\ref{JDP:one}), the
interjump size $(X_i(T_{j}^{-})-X_i(T_{j-1}^{+}))$ follows a normal
distribution of mean $\mu_i(T_{j}-T_{j-1})$ and standard deviation
$\sigma_i\sqrt{T_{j}-T_{j-1}}$. We get
\begin{eqnarray*}
X_i(T_{j}^{-})&\sim&X_i(T_{j-1}^{+})+\mu_i(T_{j}-T_{j-1})+
\sigma_{i}N(0,T_{j}-T_{j-1})\\
&\sim&X_i(T_{j-1}^{+})+\mu_i(T_{j}-T_{j-1})+
\sum_{k=1}^{N_{\mathrm{firm}}}\sigma_{ik}N(0,T_{j}-T_{j-1}),
\end{eqnarray*}
where the initial state is $X_i(0)=X_i(T_{0}^{+})$.
  \item If jump occurs, we simulate the jump-size by a normal distribution
  or another distribution when appropriate, and compute the postjump value:
\[
  X_i(T_{j}^{+})=X_i(T_{j}^{-})+Z_i(T_{j}).
\]
\end{enumerate}

This completes the procedure for generating beforejump and postjump
values $X_i(T_{j}^{-})$ and $X_i(T_{j}^{+})$. As before,
$j=1,\cdots,M$ where $M$ is the total number of jumps for all the
firms. We compute $P_{ij}$ according to Eq. (\ref{BM:default}). To
recur the first passage time density (FPTD) $f_i(t)$, we have to
consider three possible cases that may occur for each non-default
firm $i$:
\begin{enumerate}
  \item \textbf{First passage happens inside the interval.} We know that if
$X_i(T_{j-1}^{+})>D_i(T_{j-1})$ and $X_i(T_{j}^{-})<D_i(T_{j})$,
then the first passage happened in the time interval
$[T_{j-1},T_{j}]$. To evaluate when the first passage happened, we
introduce a new viable $b_{ij}$ as
$b_{ij}=\frac{T_{j}-T_{j-1}}{1-P_{ij}}$. We generate several
correlated uniform numbers $Y_i$ by using the SOU method as
described in Section \ref{subsection:SOU}, then compute
$s_i=b_{ij}Y_i+T_{j-1}$. If $s_i$ belongs to interval
$[T_{j-1},T_{j}]$, then the first passage time occurred in this
interval. We set \texttt{IsDefault}$(i)=1$ to indicate firm $i$ has
defaulted and compute the conditional boundary crossing density
$g_{ij}(s_i)$ according to Eq. (\ref{FPTD:condition}). To get the
density for the entire interval $[0,T]$, we use
$\widehat{f}_{i,n}(t)=\left(\frac{T_{j}-T_{j-1}}{1-P_{ij}}\right)g_{ij}(s_i)*K(h_{opt},t-s_i)$,
where $n$ is the iteration number of the Monte Carlo cycle.
  \item \textbf{First passage does not happen in this interval.}
If $s_i$ doesn't belong to interval $[T_{j-1},T_{j}]$, then the
first passage time has not yet occurred in this interval.
  \item \textbf{First passage happens at the right boundary of the interval.} If
$X_i(T_{j}^{+})<D_i(T_{j})$ and $X_i(T_{j}^{-})>D_i(T_{j})$ (see Eq.
(\ref{index_first_jump})), then $T_{I_i}$ is the first passage time
and $I_i=j$, we evaluate the density function using kernel function
$\widehat{f}_{i,n}(t)=K(h_{opt},t-T_{I_i})$, and set
\texttt{IsDefault}$(i)=1$.
\end{enumerate}

Next, we increase $j$ and examine the next interval and analyze the
above three cases for each non-default firm again. After running $N$
times Monte Carlo cycle, we get the FPTD of firm $i$ as
$\widehat{f}_{i}(t)=\frac{1}{N}\sum_{n=1}^{N}\widehat{f}_{i,n}(t)$.

\subsection{Model calibration}
\label{Calibaration}

We need to calibrate the developed model, in other words, to
numerically choose or optimize the parameters, such as drift,
volatility and jumps to fit the most liquid market data. This can be
done by applying the least-square method, minimizing the root mean
square error (\textit{rmse}) given by:
\[
  rmse=\sqrt{\sum_{\mathrm{derivatives}}
    \frac{(\mathrm{Market\;price}-\mathrm{Model\;price})^2}{\mathrm{Number\;of\;derivatives}}.}
  \label{Eq:calibaration}
\]
\citet{Luciano:2005} have used a set of European call options
$C(k,T)$ as their model price to calibrate their model parameters.

However, as demonstrated in Section \ref{Application}, for a number
of practically interesting cases, there is no option value that can
be used to calibrate our model, so we have to use the historical
default data to optimize the parameters in the model. As mentioned
in Sections \ref{subsection:estimation} and \ref{subsection:UNIF},
after Monte Carlo simulation we obtain the estimated density
$\widehat{f}_{i}(t)$ by using the kernel estimator method. The
cumulative default rates for firm $i$ in our model is defined as,
\begin{equation}
  P_i(t)=\int_{0}^{t}\widehat{f}_{i}(\tau)d\tau.
\end{equation}

Then we minimize the difference between our model and historical
default data $\widetilde{A}_i(t)$ to obtain the optimized parameters
in the model (such as $\sigma_{ij}$, arrival intensity $\lambda$ in
Eq. (\ref{JDP:one})):
\begin{equation}
  \mathrm{argmin}\left(\sum_{i}\sqrt{\sum_{t_j}\left(
    \frac{P_{i}(t_j)-\widetilde{A}_i(t_j)}{t_j}\right)^2}\right).
  \label{Eq:calibration:default}
\end{equation}

\section{Applications and discussion}
\label{Application}

In this section, we demonstrate the developed model at work for
analyzing the default events of multiple correlated firms via a set
of historical default data.

\subsection{Density function and default rate}
\label{application:one:firm}

First, for completeness, let us consider a set of historical default
data of differently rated firms as presented by
\citet{Zhou:2001:corr}. Our first task is to describe the first
passage time density functions and default rates of these firms.

Since there is no option value that can be used, we will employ
Eq.(\ref{Eq:calibration:default}) to optimize the parameters in our
model. For convenience, we reduce the number of optimizing
parameters by:
\begin{enumerate}
  \item Setting $X(0)=2$ and $\ln(\kappa)=0$.
  \item Setting the growth rate $\gamma$ of debt value equivalent to the
growth rate $\mu$ of the firm's value \citep{Zhou:2001:corr}, so the
default of firm is non-sensitive to $\mu$. In our computations, we
set $\mu=-0.001$.
  \item The interjump times $(T_j-T_{j-1})$ satisfy an exponential
  distribution with mean value equals to 1.
  \item The arrival rate for jumps satisfies the Poisson distribution with
intensity parameter $\lambda$, where the jump size is a normal
distribution $Z_t\sim N(\mu_{Z},\sigma_{Z})$.
\end{enumerate}

As a result, we only need to optimize $\sigma$, $\lambda$,
$\mu_{Z}$, $\sigma_{Z}$ for each firm. This is done by minimizing
the differences between our simulated default rates and historical
data. Moreover, as mentioned above, we will use the same arrival
rate $\lambda$ and distribution of $(T_j-T_{j-1})$ for differently
rated firms, so we first optimize four parameters for, e.g., the
A-rated firm, and then set the parameter $\lambda$ of other three
firms the same as A's.

The minimization was performed by the using quasi-Newton procedure
implemented as a Scilab program. The optimized parameters for each
firm are described in Table \ref{Table:param:one}.

\begin{table}[htbp]
  \centering
  \caption{Optimized parameters for differently rated firms by using the UNIF
method. The optimization was performed by using the quasi-Newton
procedure implemented as a Scilab program. In each step of the
optimization, we choose the Monte Carlo runs $N=50,000$.}
  \label{Table:param:one}
  \begin{tabular}{lcccc}
    \hline
    & $\sigma$ & $\lambda$ & $\mu_{Z}$ & $\sigma_{Z}$\\
    \hline
    A   & 0.0900 & 0.1000 & -0.2000 & 0.5000 \\
    Baa & 0.0894 & 0.1000 & -0.2960 & 0.6039 \\
    Ba  & 0.1587 & 0.1000 & -0.5515 & 1.6412 \\
    B   & 0.4500 & 0.1000 & -0.8000 & 1.5000 \\
    \hline
  \end{tabular}
\end{table}

\begin{figure}[htbp]
  \centering
  \includegraphics[width=12cm]{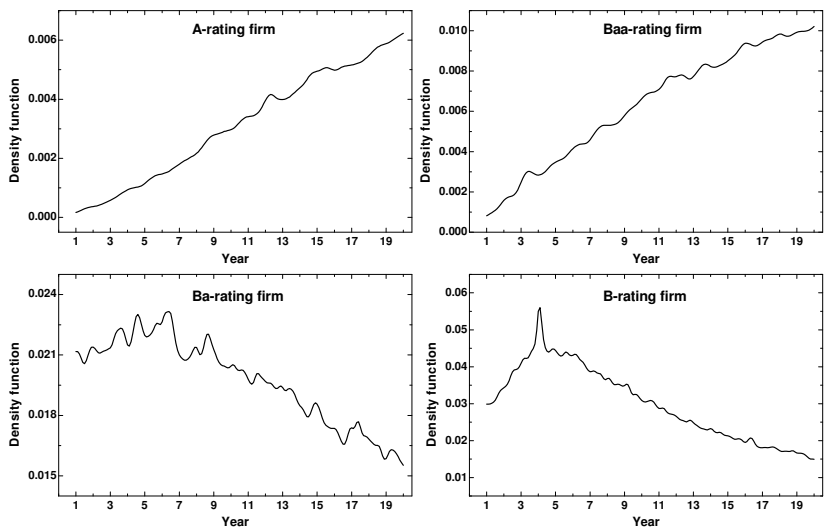}\vspace*{-1ex}
  \caption{Estimated density function for differently rated firms. All the
simulations were performed with Monte Carlo runs $N=500,000$.}
  \label{Fig:density:one}
\end{figure}

\begin{figure}[hbtp]
  \centering
  \includegraphics[width=12cm]{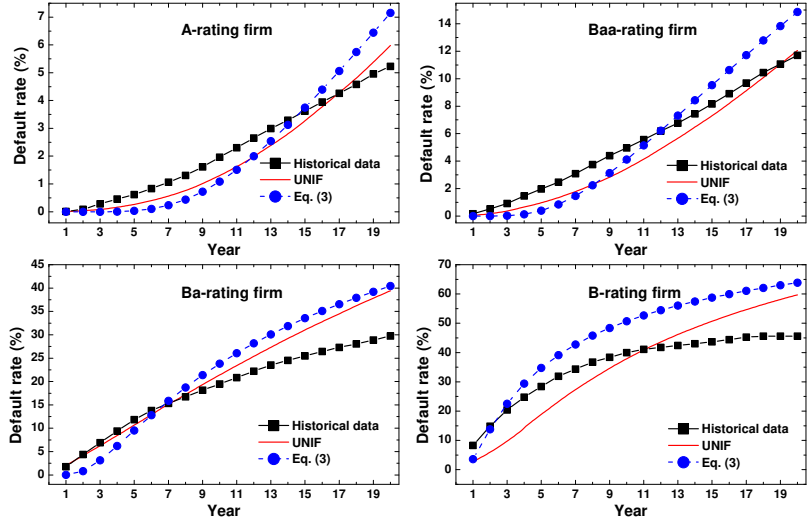}\vspace*{-1ex}
  \caption{Historical (squares), theoretical (circles) and simulated (line)
cumulative default rates for differently rated firms. All the
simulations were performed with Monte Carlo runs $N=500,000$.}
  \label{Fig:default:one}
\end{figure}

By using these optimized parameters, we carried out the final
simulation with Monte Carlo runs $N=500,000$. The estimated first
passage time density function of these four firms are shown in Fig.
\ref{Fig:density:one}. The simulated cumulative default rates (line)
together with historical data (squares) are given in Fig.
\ref{Fig:default:one}. The theoretical data denoted as circles in
Fig. \ref{Fig:default:one} were computed by using Eq.
(\ref{default:zhou:model}) where $Z_i$ were evaluated in
\citep{Zhou:2001:corr} as 8.06, 6.46, 3.73 and 2.10 for A-, Baa-,
Ba- and B-rated firms, respectively. In Table \ref{Tab:hopt}, we
give the optimal bandwidth and parameters $\alpha$, $\beta$ for the
true density estimate.

\begin{table}[hbtp]
  \centering
  \caption{The optimal bandwidth $h_{opt}$, parameters $\alpha$,
$\beta$ for the true density estimate of differently rated firms.
All the simulations were performed with Monte Carlo runs
$N=500,000$.}
  \label{Tab:hopt}
  \begin{tabular}{lccc}
    \hline
    & $\alpha$ & $\beta$ & Optimal bandwidth\\
    \hline
    A   & 0.206699 & 3 & 0.655522 \\
    Baa & 0.219790 & 3 & 0.537277 \\
    Ba  & 0.252318 & 3 & 0.382729 \\
    B   & 0.327753 & 3 & 0.264402 \\
    \hline
  \end{tabular}
\end{table}

Based on these results, we conclude that:
\begin{enumerate}
  \item Simulations give similar or better results to the analytical
  results predicted by Eq. (\ref{default:zhou:model}).
  \item A- and Baa-rated firms have a smaller Brownian motion part. Their
  parameters $\sigma$ are much smaller than those of Ba- and B-rated firms.
  \item The optimized parameters $\sigma$ of A- and Baa-rated firms
  are similar, but the jump parts $(\mu_Z,\sigma_Z)$ are different,
  which explains their different cumulative default rates and
  density functions. Indeed, Baa-rated firm may encounter more severe
  economic hazard (large jump-size) than A-rated firm.
  \item As for Ba- and B-rated firms, except for the large $\sigma$, both of
  them have large $\mu_Z$ and especially large $\sigma_Z$, which
  indicate that the loss due to sudden economic hazard may
  fluctuate a lot for these firms. Hence, the large $\sigma$,
  $\mu_Z$ and $\sigma_Z$ account for their high default rates and low
  credit qualities.
  \item From Fig. \ref{Fig:density:one}, we can conclude that the density
  functions of A- and Baa-rated firms still have the trend to
  increase, which means the default rates of A- and Baa-rated firms
  may increase little faster in future. As for Ba- and
  B-rated firms, their density functions have decreased,
  so their default rates may increase very slowly or be kept at a
  constant level. Mathematically speaking, the cumulative default rates of
  A- and Baa-rated firms are convex function, while the cumulative
  default rates of Ba- and B-rated firms are concave.
\end{enumerate}

\subsection{Correlated default}
\label{application:multi:firm}

Our final example concerns with the default correlation of two
firms. If we do not include jumps in the model, the default
correlation can be calculated by using Eq.
(\ref{default:zhou:model}), (\ref{Eq:Por}) and (\ref{Eq:corr2}). In
Tables \ref{Simulate:Corr01}-\ref{Simulate:Corr10} we present
comparisons of our results with those based on closed form solutions
provided by \citet{Zhou:2001:corr} with $\rho=0.4$.

\begin{table}[htbp]
  \centering
  \caption{One year default correlations (\%). All the simulations are
performed with the Monte Carlo runs $N=500,000$}
  \label{Simulate:Corr01}
  \begin{tabular}{lcccccccc}
    \hline
      & \multicolumn{4}{c}{UNIF} & \multicolumn{4}{c}{\citet{Zhou:2001:corr}} \\
    \hline
        & A     & Baa  & Ba    & B     & A    & Baa  & Ba   & B     \\
    A   & -0.01 &      &       &       & 0.00 &      &      &       \\
    Baa & -0.02 & 3.69 &       &       & 0.00 & 0.00 &      &       \\
    Ba  &  2.37 & 4.95 & 19.75 &       & 0.00 & 0.01 & 1.32 &       \\
    B   &  2.80 & 6.63 & 22.57 & 26.40 & 0.00 & 0.00 & 2.47 & 12.46 \\
    \hline
  \end{tabular}
\end{table}

\begin{table}[htbp]
  \centering
  \caption{Two year default correlations (\%). All the simulations are
performed with the Monte Carlo runs $N=500,000$}
  \label{Simulate:Corr02}
  \begin{tabular}{lcccccccc}
    \hline
    & \multicolumn{4}{c}{UNIF} & \multicolumn{4}{c}{\citet{Zhou:2001:corr}} \\
    \hline
        & A    & Baa  & Ba    & B     & A    & Baa  & Ba   & B     \\
    A   & 2.35 &      &       &       & 0.02 &      &      &       \\
    Baa & 2.32 & 4.25 &       &       & 0.05 & 0.25 &      &       \\
    Ba  & 4.17 & 7.17 & 20.28 &       & 0.05 & 0.63 & 6.96 &       \\
    B   & 4.73 & 8.23 & 23.99 & 29.00 & 0.02 & 0.41 & 9.24 & 19.61 \\
    \hline
  \end{tabular}
\end{table}

\begin{table}[htbp]
  \centering
  \caption{Five year default correlations (\%). All the simulations are
performed with the Monte Carlo runs $N=500,000$}
  \label{Simulate:Corr05}
  \begin{tabular}{lcccccccc}
    \hline
    & \multicolumn{4}{c}{UNIF} & \multicolumn{4}{c}{\citet{Zhou:2001:corr}} \\
    \hline
        & A    & Baa   & Ba    & B     & A    & Baa  & Ba    & B     \\
    A   & 6.45 &       &       &       & 1.65 &      &       &       \\
    Baa & 6.71 & 9.24  &       &       & 2.60 & 5.01 &       &       \\
    Ba  & 7.29 & 10.88 & 22.91 &       & 2.74 & 7.20 & 17.56 &       \\
    B   & 6.77 & 10.93 & 22.97 & 27.93 & 1.88 & 5.67 & 18.43 & 24.01 \\
    \hline
  \end{tabular}
\end{table}

\begin{table}[htbp]
  \centering
  \caption{Ten year default correlations (\%). All the simulations are
performed with the Monte Carlo runs $N=500,000$}
  \label{Simulate:Corr10}
  \begin{tabular}{lcccccccc}
    \hline
    & \multicolumn{4}{c}{UNIF} & \multicolumn{4}{c}{\citet{Zhou:2001:corr}} \\
    \hline
        & A     & Baa   & Ba    & B     & A    & Baa   & Ba    & B     \\
    A   & 8.79  &       &       &       & 7.75 &       &       &       \\
    Baa & 10.51 & 13.80 &       &       & 9.63 & 13.12 &       &       \\
    Ba  & 9.87  & 14.23 & 22.50 &       & 9.48 & 14.98 & 22.51 &       \\
    B   & 8.50  & 12.54 & 20.49 & 24.98 & 7.21 & 12.28 & 21.80 & 24.37 \\
    \hline
  \end{tabular}
\end{table}

Next, let us consider the default correlations under the
multivariate jump-diffusion processes. We use the following
conditions in our multivariate UNIF method:
\begin{enumerate}
  \item Setting $X(0)=2$ and $\ln(\kappa)=0$ for all firms.
  \item Setting $\gamma=\mu$ and $\mu=-0.001$ for all firms.
  \item Since we are considering two correlated firms, we choose $\sigma$ as,
\begin{equation}
  \sigma=\left[
  \begin{tabular}{cc}
    $\sigma_{11}$ & $\sigma_{12}$\\
    $\sigma_{21}$ & $\sigma_{22}$
  \end{tabular}\right],
\end{equation}
where $\sigma\sigma^\top=H_0$ such that,
\[
  \sigma\sigma^\top=H_0=\left[
  \begin{tabular}{cc}
    $\sigma_1^2$ & $\rho_{12}\sigma_1\sigma_2$ \\
    $\rho_{12}\sigma_1\sigma_2$ & $\sigma_2^2$
  \end{tabular}\right],
\]
and
\begin{equation}
  \left\{
  \begin{tabular}{l}
    $\sigma_1^2=\sigma_{11}^2+\sigma_{12}^2$,\\
    $\sigma_2^2=\sigma_{21}^2+\sigma_{22}^2$,\\
    $\rho_{12}=\displaystyle\frac{\sigma_{11}\sigma_{21}+\sigma_{12}\sigma_{22}}{\sigma_1\sigma_2}$.
  \end{tabular}\right.
  \label{Eq:Brownian:corr}
\end{equation}
In Eq. (\ref{Eq:Brownian:corr}), $\rho_{12}$ reflects the
correlation of diffusion parts of the state vectors of the two
firms. In order to compare with the standard Brownian motion and to
evaluate the default correlations between different firms, we set
all the $\rho_{12}=0.4$ as in \citet{Zhou:2001:corr}. Furthermore,
we use the optimized $\sigma_1$ and $\sigma_2$ in Table
\ref{Table:param:one} for firm 1 and 2, respectively. Assuming
$\sigma_{12}=0$, we get,
\[
  \left\{
  \begin{tabular}{l}
    $\sigma_{11}=\sigma_1$,\\
    $\sigma_{12}=0$,\\
    $\sigma_{21}=\rho_{12}\sigma_2$,\\
    $\sigma_{22}=\sqrt{1-\rho_{12}^2}\sigma_2$.
  \end{tabular}\right.
\]
  \item The arrival rate for jumps satisfies the Poisson distribution with
intensity parameter $\lambda=0.1$ for all firms. The jump size is a
normal distribution $Z_t\sim N(\mu_{Z_i},\sigma_{Z_i})$, where
$\mu_{Z_i}$ and $\sigma_{Z_i}$ can be different for different firms
to reflect specifics of the jump process for each firm. We adopt the
optimized parameters given in Table \ref{Table:param:one}.
  \item As before, we generate the same interjump times $(T_j-T_{j-1})$ that
satisfy an exponential distribution with mean value equals to 1 for
each two firms.
\end{enumerate}

We carry out the UNIF method to evaluate the default correlations
via the following formula:
\begin{equation}
  \rho_{12}(t)=\frac{1}{N}\sum_{n=1}^{N}
  \frac{P_{12,n}(t)-P_{1,n}(t)P_{2,n}(t)}{\sqrt{P_{1,n}(t)(1-P_{1,n}(t))P_{2,n}(t)(1-P_{2,n}(t))}},
  \label{Eq:simulate:corr}
\end{equation}
where $P_{12,n}(t)$ is the probability of joint default for firms 1
and 2 in each Monte Carlo cycle, $P_{1,n}(t)$ and $P_{2,n}(t)$ are
the cumulative default rates of firm 1 and 2, respectively, in each
Monte Carlo cycle.

The simulated default correlations for one-, two-, five- and
ten-years are given in Table
\ref{Simulate:Corr01}-\ref{Simulate:Corr10}. All the simulations
were performed with the Monte Carlo runs $N=500,000$. Comparing
those simulated default correlations with the theoretical data for
standard Brownian motions, we can conclude that
\begin{enumerate}
  \item Similarly to conclusions of \citet{Zhou:2001:corr}, the default
  correlations of same rated firms are usually large compared to
  differently rated firms. Furthermore, the default correlations tend
  to increase over long horizons and may converge to a stable value.
  \item In our simulations, the one year default correlations of
  (A,A) and (A,Baa) are negative. This is because they seldom
  default jointly during one year. Note, however, that the default
  correlations of other firms are positive and usually larger than the
  results in \citet{Zhou:2001:corr}.
  \item For two and five years, the default correlations of
  different firms increase. This can be explained by the fact that
  their individual first passage time density functions increase
  during these time horizon, hence the probability of joint default
  increases.
  \item As for ten year default correlations, our simulated results
  are almost identical to the theoretical data for standard Brownian
  motions. The differences are that the default correlations of
  (Ba,Ba), (Ba,B) and (B,B) decrease from the fifth year to tenth year
  in our simulations. The reason is that the first passage time
  density function of Ba- and B-rated firms begin to decrease from
  the fifth year, hence the probability of joint default may
  increase slowly.
\end{enumerate}

\section{Conclusion}
\label{Conclusion}

In this contribution, we have analyzed the credit risk problems of
multiple correlated firms in the structural model framework, where
we incorporated jumps to reflect the external shocks or other
unpredicted events. By combining the fast Monte-Carlo method for
one-dimensional jump-diffusion processes and the generation of
correlated multidimensional variates, we have developed a fast
Monte-Carlo type procedure for the analysis of multivariate and
correlated jump-diffusion processes. The developed approach
generalizes previously discussed non-correlated jump-diffusion cases
for multivariate and correlated jump-diffusion processes. Finally,
we have applied the developed technique to analyze the default
events of multiple correlated firms via a set of historical default
data. The developed methodology provides an efficient computational
technique that is applicable in other areas of credit risk and
pricing options.

\section*{Acknowledgments}
\label{Acknowledgements}

This work was supported by NSERC.




\end{document}